\documentclass[manuscript]{acmart}
\AtBeginDocument{%
  }

\usepackage{graphicx} 
\usepackage{subfig}
\usepackage{hyperref}
\usepackage{multirow}

\copyrightyear{2022}
\acmYear{2022}
\setcopyright{rightsretained}
\acmConference[FAccT '22]{2022 ACM Conference on Fairness, Accountability, and Transparency}{June 21--24, 2022}{Seoul, Republic of Korea}
\acmBooktitle{2022 ACM Conference on Fairness, Accountability, and Transparency (FAccT '22), June 21--24, 2022, Seoul, Republic of Korea}
\acmDOI{10.1145/3531146.3533157}
\acmISBN{978-1-4503-9352-2/22/06}

\begin{document}

\title{The Forgotten Margins of AI Ethics}
\author{Abeba Birhane}
\email{abeba@mozillafoundation.org}
\orcid{0000-0001-6319-7937}
\affiliation{%
  \institution{Mozilla Foundation \& School of Computer Science, University College Dublin}
  \streetaddress{Belfied}
  \city{Dublin}
  \country{Ireland}
}
\author{Elayne Ruane}
\email{elayne.ruane@ucdconnect.ie}
\orcid{0000-0001-7344-9763}
\affiliation{%
  \institution{SFI Lero \& School of Computer Science, University College Dublin}
  \streetaddress{Belfied}
  \city{Dublin}
  \country{Ireland}
}
\author{Thomas Laurent}
\email{thomas.laurent@ucd.ie}
\orcid{0000-0002-0953-774X}
\affiliation{%
  \institution{SFI Lero \& School of Computer Science, University College Dublin}
  \streetaddress{Belfied}
  \city{Dublin}
  \country{Ireland}
}
\author{Matthew S. Brown}
\email{msb027@bucknell.edu}
\orcid{}
\affiliation{%
  \institution{Department of Computer Science, Bucknell University}
  \streetaddress{}
  \city{}
  \state{}
  \country{USA}
  \postcode{}
}
\author{Johnathan Flowers}
\email{j.charlesflowers@gmail.com}
\orcid{}
\affiliation{%
  \institution{Department of Philosophy \& Religion, American University}
  \streetaddress{}
  \city{}
  \state{}
  \country{USA}
  \postcode{}
}
\author{Anthony Ventresque}
\email{anthony.ventresque@ucd.ie}
\orcid{0000-0003-2064-1238}
\affiliation{%
  \institution{SFI Lero \& School of Computer Science, University College Dublin}
  \streetaddress{Belfied}
  \city{Dublin}
  \country{Ireland}
}
\author{Christopher L. Dancy}
\email{cdancy@psu.edu}
\orcid{0000-0003-2119-496X}
\affiliation{%
  \institution{Dept of Industrial and Manufacturing Engineering \& Dept of Computer Science and Engineering, The Pennsylvania State University}
  \streetaddress{}
  \city{University Park}
  \state{PA}
  \country{USA}
  \postcode{}
}
\renewcommand{\shortauthors}{Birhane, et al.}

\begin{abstract}
How has recent AI Ethics literature addressed topics such as fairness and justice in the context of continued social and structural power asymmetries? 
We trace both the historical roots and current landmark work that have been shaping the field and categorize these works under three broad umbrellas: (i) those grounded in Western canonical philosophy, (ii) mathematical and statistical methods, and (iii) those emerging from critical data/algorithm/information studies.  
We also survey the field and explore emerging trends by examining the rapidly growing body of literature that falls under the broad umbrella of AI Ethics. To that end, we read and annotated peer-reviewed papers published over the past four years in two premier conferences: FAccT and AIES. We organize the literature based on an annotation scheme we developed according to three main dimensions: whether the paper deals with concrete applications, use-cases, and/or people's lived experience; to what extent it addresses harmed, threatened, or otherwise marginalized groups; and if so, whether it explicitly names such groups. 
We note that although the goals of the majority of FAccT and AIES papers were often commendable, their consideration of the negative impacts of AI on traditionally marginalized groups remained shallow. Taken together, our conceptual analysis and the data from annotated papers indicate that the field would benefit from an increased focus on ethical analysis grounded in concrete use-cases, people's experiences, and applications as well as from approaches that are sensitive to structural and historical power asymmetries. 
\end{abstract}


\maketitle

\section{Introduction}
\label{introduction}

The urgency and profound importance of fairness, justice, and ethics in Artificial Intelligence (AI) has been made evident by grave failures, limitations, harms, threats, and inaccuracies emanating from algorithmic systems deployed in various domains from criminal justice systems, and education, to medicine. These have been brought forth by landmark studies~\cite{angwin2016machine, buolamwini2018gender,obermeyer2019dissecting} and foundational books~\cite{noble2018algorithms, o2016weapons, eubanks2018automating, pasquale2015black, gandy2021panoptic, costanza2020design, d2020data, zuboff2019age}. Furthermore, the critical importance of the topic is marked by newly founded conferences exclusively dedicated to AI Ethics (e.g.\ AIES and FAccT), the fast-growing adoption of ethics into syllabi in computational departments~\cite{fiesler2020we}, newly introduced requirements for the inclusion of broader impacts statements for AI and Machine Learning (ML) papers submitted in premier AI conferences such as NeurIPS\footnote{\url{https://nips.cc/Conferences/2020/CallForPapers}}, increased attention to policy and regulation of AI~\cite{jobin2019global}, and increasing interest in ethics boards and research teams dedicated to ethics in major tech corporations. 

This urgency and heightened attention on fairness, justice, and ethics is warranted. AI research, tools, and models --- especially those integrated into the social sphere --- are not purely technical systems that remain in the lab but socio-technical systems that are often deployed into the social world with potential and real impact on individual people, communities, and society at large. From discriminatory policing~\cite{scannell2019not,browning2020stop,lum2016predict} and discriminatory medical care allocations~\cite{obermeyer2019dissecting,vyas2020hidden,powe2020black}, to discriminatory targeted ads~\cite{speicher2018potential,ali2019discrimination,kuhn2013gender}, algorithmic systems present real harm to the most marginalized groups of society and furthermore pose a threat to human welfare, freedom, and privacy. 

The broadly construed field of AI Ethics has been crucial in bringing the societal implications of algorithmic systems to the fore. It is important to note that critical analysis of computing has existed as long as the computing and AI fields themselves~\cite{weizenbaum1976computer, dreyfus1972computers, winograd1986understanding, agre1994surveillance}. Nonetheless, over the past few years a robust body of work has established the field of AI Ethics as a foundational and essential endeavour within the broad AI and society research. 
Work on fairness, accountability, transparency, explainability, bias, equity, and justice are generally conceived under this broad umbrella concept of AI Ethics. 

With algorithmic bias, harm, discrimination, and injustice in the spotlight, the broadly construed field of AI Ethics is growing rapidly, producing a large and diverse body of work over the past number of years. This growing field, far from representing a unified entity, is marked by plural, overlapping, competing, and at times contradictory values, objectives, methods, perspectives, and goals. Yet, despite this plurality, the emerging field occupies a critical role in setting standards and norms for AI ethics research and practice. From informing policy and establishing ethical standards, to (directly or indirectly) impacting funding and influencing research agendas, as well as shaping the general discourse of AI Ethics, this fast expanding field impacts the current and future direction of fairness, justice, and ethics research and practice in AI. 

Historically, within the Western context, ``classic'' ethics, as a branch of philosophy has mainly been speculative and theoretical where queries of AI ethics emanating from them can veer towards hypothetical questions such as ``threats from super-intelligence''. While theoretical rigour is crucial, it also poses a risk where questions of AI ethics could get lost in abstract speculation at the expense of currently existing harms to actual people. As such, it is important to review the historical roots of the field, examine the limitations of ``classic'' ethics, understand the current state and emerging patterns of the field, explore where it might be heading, and provide recommendation towards AI ethics research and practice that is grounded in the concrete. To that end, in this paper we: 

\begin{itemize}
    \item Trace historical roots of current AI ethics scholarship and provide a taxonomy of the field in terms of three overlapping categories: computational, philosophical, and critical.
    \item Carry an extensive conceptual analysis, and argue that critical approaches to AI ethics hold the most fruitful and promising way forward in terms of benefiting the most disproportionately impacted groups at the margins of society.
    \item Examine research publications from the two most prominent AI Ethics venues --- FAccT and AIES --- and present emerging trends and themes. 
\end{itemize}

The rest of the paper is structured as follows. Section~\ref{sect:history_roots} provides a historical analysis outlining the existential tension between ethical perspectives that put emphasis on theoretical and abstract conceptualizations on the one hand, and those that work from concrete particulars on the other; highlighting the limitations of abstraction in the context of AI ethics. Section~\ref{sect:the_limits} outlines how critical approaches that interrogate systemic, societal, and historical oppression; challenge existing power asymmetries; and explicitly name the negative societal and ethical implications of AI systems in a given application, group, or context are likely to bring about actionable change.  Section~\ref{sect:surveying} presents the methodological details for our systematic analysis of the trends of the field through publications in the FAccT and AIES conferences. This is followed by a quantitative summary of the the main patterns of current publications in the two conferences' proceedings in Section~\ref{sect:quant}. Section~\ref{sect:qual} takes an in-depth qualitative look at how one of the canonical works --- the COMPAS algorithms --- has been covered in the literature. Section~\ref{sect:conclusion} concludes the paper.

\section{The irremovable tension between the abstract and concrete} 
\label{sect:history_roots}

The topic of ethics, particularly within the Western context --- like other branches of philosophy --- is often a speculative (rather than a practical) endeavour~\cite{gardiner1996alterity}. As a branch of Western philosophy, the broad field of ethics enquires, assesses, and theorizes about questions of justice, virtue, fairness, good and evil, and right and wrong. Such examinations often treat questions of ethics in an abstract manner, aspiring to arrive at universal theories or principles that can be uniformly applied regardless of time, place, or context~\cite{markova2016dialogical,de2019loving,shotter2006vygotsky}. Canonical Western approaches to ethics, from deontology to consequentialism, at their core strive for such universal and generalisable theories and principles ``uncontaminated'' by a particular culture, history, or context. Underlying this aspiration is the assumption that theories and principles of ethics can actually be disentangled from contingencies and abstracted in some form devoid of context, time, and space~\cite{juarrero2000dynamics,daston2018calculation}. What’s more, this ambition for universal principles in ethics has been dominated by a particular mode of being human~\cite{Wynter03,ahmed2007phenomenology}; one that takes a straight, white ontology as a foundation and recognizes the privileged Western white male as the standard, quintessential representative of humanity~\cite{ahmed2007phenomenology}. 

Traditions like Black Feminism~\cite{h1984feminist,collins2002black}, Care Ethics~\cite{noddinge2013caring}, and the broad traditions of ethics that have emerged from Asian~\cite{li1994,ameshall2003} and African~\cite{mbiti1969african,fromrationality2020,tamale2020decolonization} scholars challenge the ontologies presumed in the ``canonical'' approach to ethics common to both the field of philosophy and the growing field of AI Ethics. These traditions are often grounded in down-to-earth problems and often strive to challenge underlying oppressive social structures and uneven power dynamics. 
Although there have been notable attempts to incorporate non-Western and feminist care ethics into the existing ethical paradigms within AI Ethics and Western philosophy~\cite{vallor2016,mohamed2020decolonial}, these attempts remain marginalized by the structure of both disciplines. This situation is further exacerbated by the paucity of available expertise at the instructor level for less commonly taught philosophies, which has led to a dangerous reification of how philosophy, and ethics by extension, should be applied~\cite{v2017taking}. Given the ways that existing AI Ethics literature builds on the circulation of existing philosophical inquiry into ethics, the reproduction of the exclusion of marginalized philosophies and systems of ethics in AI Ethics is unsurprising, as is the maintenance of the white, Western ontologies upon which they are based.

Abstraction can be, and in some domains such as mathematics and art is, a vital process to discern and highlight important features, patterns, or aspects of a subject of enquiry --- a way of seeing the forest for the trees. Additionally, clearly defining and delineating what is abstract and what is concrete is not always possible, and also  can be an undesirable binary categorization. 
Our discussion of abstract and concrete, therefore, should be seen in the broadest terms and with these qualification. While abstractions and aspirations for universal principles can be worthy endeavours, whether such an objective is useful or suitable when it comes to thinking about ethics within the realm of AI and ML, is questionable. Scholars such as Heinz von Foerster and Bernhard Pörksen~\cite{von2002understanding} have argued that: \textit{``Ethics is a matter of practice, of down-to-earth problems and not a matter of those categories and taxonomies that serve to fascinate the academic clubs and their specialists.''} Selbst et al.~\cite{selbst2019fairness} further contend that, when used to define notions of fairness and to produce fairness-aware learning algorithms, abstraction renders them ``\textit{ineffective, inaccurate, and sometimes dangerously misguided.}'' Accordingly, the authors emphasize that abstraction, which is a fundamental concept of computing, also aids to sever ML tools from the social contexts in which they are embedded. The authors identify five traps, which they collectively name \textit{the abstraction traps}, within the AI Ethics literature and the field of Computer Science in general, which result from failure to consider specific social contexts. They  further note that: ``\textit{To treat fairness and justice as terms that have meaningful application to technology separate from a social context is therefore to make a category error [...] an abstraction error.}'' On a similar note, Crawford contends that abstraction mainly serves the already powerful in society \textit{``The structures of power at the intersection of technology, capital and governance are well served by this narrow, abstracted analysis.''}~\cite{crawford2021atlas} 

When algorithmic systems fail --- and they often do --- such failures result in concrete negative outcomes such as job loss or exclusion from opportunities. Landmark work surrounding algorithmic decision-making in recent years has brought to light issues of discrimination, bias, and harm, demonstrating harms and injustices in a manner that is grounded in concrete practices, specific individuals, and groups. Computer Vision systems deployed in the social world suffer from disproportionate inaccuracies when it comes to recognizing faces of Black women compared to that of white men~\cite{buolamwini2018gender}; when social welfare decision-making processes are automated, the most vulnerable in society pay the heaviest price~\cite{eubanks2018automating}; and search engines continually perpetuate negative stereotypes about women of colour~\cite{noble2018algorithms}; to mention but a few examples. In short, when problems that arise from algorithmic decision-making are brought forth, we find that those disproportionately harmed or discriminated against are individuals and communities already at the margins of society. 

This connection between AI (and AI-adjacent) systems and negative societal impact is not a new phenomenon. Indeed, one of the earlier police computing systems in the USA was used starting in the 1960s by the Kansas City police department to (among other tasks) analyze and allocate resources for operations and planning, which included police officer deployment~\cite{mcilwain2019black}. By using not only attributes such as the race of individuals arrested, but also geographically linked data (such as number of crimes within a given area), the system created an increased surveillance and policing state that was particularly devastating for the Black population of Kansas City (see~\cite{rothstein2017color}, and~\cite{taylor2019race}, for historical discussions of the laws and practices that led to geographic segregation by race that continues to be a feature of USA neighborhoods). This is not a particularly surprising outcome if one looks beyond strictly computing systems to consider how various technologies have been used across space and time to surveil and police Black people in the USA~\cite{browne2015dark}. 
Similarly, as Weizenbaum~\cite {weizenbaum1976computer} pointed out in 1976, the computer has from the beginning been a fundamentally conservative force which solidified existing power --- in place of fundamental social changes. Using the automation of the US banking system (which Weizenbaum helped develop), as a primary example, he argued that the computer renders technical solutions that allow existing power hierarchies to remain intact. 

Tracking the relation of current AI systems to historical systems and technologies (and, indeed, social systems) allows us to more correctly contextualize AI systems (and by extension considerations of AI Ethics) within the existing milieu of socio-technical systems. This gives us an opportunity to account for the fact that \textit{``social norms, ideologies, and practices are a constitutive part of technical design''}~\cite [p.41]{benjamin2019race}. Benjamin~\cite{benjamin2019race} makes a useful connection to the ``Jim Crow'' era in the United States in coining the term the ``New Jim Code''. She argues that we might consider race and racialization as, itself, a technology to \textit{``sort, organize, and design a social structure}''~\cite[p.91]{benjamin2019race}. Thus, AI Ethics work addressing fairness, bias, and discrimination must consider how a drive for efficiency, and even inclusion, might not actually move towards a ``social good'', but instead accelerate and strengthen technologies that further entrench conservative social structures and institutions.  
Examining the broader social, cultural, and economic context as well as explicitly emphasizing concrete and potential harms is crucial when approaching questions of AI ethics, social impact of AI, and justice. This focus on broader social factors and explicit emphasis on harms allows for the field to move forward by first understanding the precise conditions and realities on the ground, acknowledging current and historical injustices and power imbalances, envisioning an alternative future, and working toward it in a manner that brings about material change to the most negatively impacted by technology. 

\subsection{The limits of fairness}
\label{sect:the_limits}

One of the ways ethics is studied in the Western AI ethics space is through the concept of algorithmic fairness. 
While attempts to formalise the concept mathematically and operationalize quantitative measures of fairness may reduce vagueness, such approaches often treat fairness in abstract terms. It is also important to note that algorithmic fairness is a slippery notion that is understood in many different ways, typically with mutually incompatible conceptions~\cite{friedler2016possibility}. The treatment of concepts such as fairness and bias in abstract terms is frequently linked to the notion of neutrality and objectivity, the idea that there exists a ``purely objective'' dataset or a ``neutral'' representation of people, groups, and the social world independent of the observer/modeller~\cite{birhane2021algorithmic,benjamin2019race}. This is often subsequently followed by simplistic, reductive, and shallow efforts to ``fix'' such problems such as ``debiasing'' datasets --- a practice that may result in more harm as it gives the illusion that the problem is solved, disincentivizing efforts that 
engage with the roots causes (often structural) of the problem~\cite{gonen2019lipstick}. At a minimum, ``debiasing'' approaches place problems on datasets and away from contingent wider societal and structural factors.      

The discussion of ``fairness'' as an abstract mathematical or philosophical concept makes actual consequences from failures of AI systems a distant statistical or conceptual problem, effectively allowing us to overlook impacted individuals behind each data point~\cite{d2020data,raji2020discomfort}. This is where analysis of power asymmetries proves critical. Fairness in abstract terms as a statistic or a number in a table might be something the most privileged in society, who are not impacted by AI systems, find convenient. 
However, for those who carry the brunt of algorithmic failures, fairness is inseparable from their daily lives, and very concrete indeed. 

An approach to ethics that ignores historical, societal, and structural issues omits factors of critical importance.
In \textit{``Don't ask if artificial intelligence is good or fair, ask how it shifts power''}, Kalluri~\cite{kalluri2020don} argues that genuine efforts to ethical thinking are those that bring about concrete material change to the conditions of the most marginalized, those approaches that shift power from the most to the least powerful in society. Without critical considerations of structural systems of power and oppression serving a central role in the ethical considerations of AI systems, researchers, designers, developers, and policy-makers risk equating processes and outcomes that are \textit{fair} with those that are \textit{equitable} and \textit{just}. Engaging with the dangers of scale that AI-related algorithmic systems pose requires understanding and accounting for power structures underlying them. This is especially true where AI systems are adapted to work within said ongoing systems of power and oppression to efficiently scale the effects of those systems on marginalized populations~\cite{eubanks2018automating,noble2018algorithms}.

Despite the recent calls against de-coupling socio-cultural considerations from the (socio-)technical AI systems built and to sustain a critical perspective that take systems of power into consideration~\cite{dancy21blackness, selbst2019fairness, hampton21blackfem}, the lack of critical perspectives largely persists. Considerations of ethics and fairness in AI systems must make structural systems of power and oppression central to the design and development of such systems and resist the overly (techno)optimistic perspectives that assume context-less fairness will result in equitable outcomes. Furthermore, these considerations must directly contend with the possibility that AI systems may be inequitable and oppressive by their very application within particular contexts. The next section presents the current state of the field in relation to such critical analysis through the study of papers presented at two premiere AI Ethics conferences.

\section{Surveying the field} 
\label{sect:surveying}
 
Our approach ``follows'' AI Ethics (as conceptualized by Western societies) through publications from two of the most prominent AI ethics venues: FAccT and AIES. 
Over the past five years, FAccT and AIES have risen to be the top two conferences where the community gathers to discuss, debate and publish state-of-the-art findings and ideas.  
By examining the publications that make up recent scholarship on AI Ethics, our review reveals the shape and contours of how ``ethics'' is conceptualized, operationalized, and researched and where it might be heading.  
Through this approach, we can come to understand the ``work'' that ethics does in the field of AI Ethics, where ethics ``goes'' insofar as how specific culturally-coded modes of understanding are circulated throughout the field, as well as the boundaries of what is and is not considered ethical. 
Following ethics in this way, therefore, provides crucial insights into the development of AI Ethics, and what the subject of legitimate AI Ethics inquiry should be. 

\subsection{Method} 
We read and annotated all papers published in the FAccT (formerly known as FAT*) and AIES conference proceedings over the past four years, from the founding of the conferences in 2018 to date (2021). This resulted in a total of 535 papers. Of those, we analysed 526 papers. For consistency, 91\% of papers were annotated by two independent annotators. Overall, the annotators strongly agreed on all measures. This is reflected by the Cohen's Kappa~\cite{cohen1960coefficient}, a common inter-rater agreement measure, values of 0.93, 0.91, and 0.92 for the concreteness, disparity, and explicitness measures respectively. We recorded the paper title, authors, year of publication, publication venue, and any author-defined keywords. For each paper, we then read and annotated the abstract, introduction, and conclusion and/or discussion section of each paper according to an annotation scheme we developed (described below).

\subsection{Annotation Scheme}
Through an iterative process, we created an annotation scheme to classify and score each paper on three dimensions. We collectively established a set of guidelines that served as a point of reference for annotators during the annotation process. 
Each paper was annotated by two independent annotators each of whom completed a three-part annotation classification and scoring process as follows:
\begin{enumerate}
    \item \textbf{Concreteness Measure.} Annotate the paper along a binary~\footnote{We recognize the limitations of binary coding of papers into either/or categories. To counterbalance these shortcomings, we also present an in depth qualitative analysis using the COMPAS coverage as a case study. Further nuanced elaboration on the problems of delineating the abstract from the concrete can be found in Section~\ref{sect:history_roots}. 
    } concreteness measure with values \textit{concrete} or \textit{abstract}. 
        \begin{itemize}
            \item A paper is labelled \textit{abstract} if it approaches concepts such as ethics, fairness, harm, risk, and bias (shortened as ethics from here on) in a manner where context such as existing technological applications and/or discussion of specific impacted groups, classes, or demographics are absent. This might include presupposition of universal principles, invoking hypotheticals such as thought experiments, suppositions, if-then statements, and/or framing of such concepts as a primarily mathematical formulation when addressing the aforementioned ethical concepts (see Section~\ref{sect:history_roots} for further discussion on abstract and concrete). 
            \item \textit{Concrete}, on the other hand, is a label applied to a paper that approaches ethical concepts in a manner that is grounded in existing applications of a certain tool or technology, is rooted in specific demographics, living people, and/or culture and context-specific socio-technical phenomena.    
        \end{itemize}   
    \item \textbf{Disparity Measure.} We used disparity measures to assess whether a paper investigates, addresses, and/or mentions the disparate impacts of an algorithmic system. We identified five dimensions defined as follows. Each paper was assigned one of the five values.   
        \begin{itemize}
            \item \textit{Strong} --- the primary focus of the paper centers around exposing and tackling (through theoretical analysis, empirical experiments, and/or critical analysis) disparate impacts, possibly towards particular groups, classes, or demographics. 
            \item \textit{Medium} --- the paper engages with disparate impacts (e.g., by addressing “protected attributes”) in a substantial manner but falls short of making such discussion or analysis its primary objective. 
            \item \textit{Weak} --- the paper mentions, perhaps in passing, disparate impact of algorithmic systems but lacks any significant engagement with the topic.
            \item \textit{Agnostic} --- the paper does not mention any impact of algorithmic systems on individuals, groups, communities, or demographics.
            \item \textit{Inadequate} (\textit{naive}, \textit{ill-considered}, \textit{ill-advised}, or \textit{anti-equity}) --- the paper may mention or even engage with the topic of disparate impact but does so in a way that shows a certain naivety of the topic or in the impact of the application of the algorithmic system, or the work in the paper may even explicitly enact disparate impacts, whether at the individual or institutional level. As the annotation scheme was developed in an iterative process, this dimension was added to the scheme at a later stage following a handful of recurring themes that emerged in the literature. Papers in this category also include those dealing with sensitive issues, for example, work proposing the use of algorithms for identifying gang-related violent crimes with no discussion of social and environmental factors or social impact and/or harm the work might cause to individuals or groups deemed to be ``gang members''.   
        \end{itemize}
    \item \textbf{Explicitness Measure.} For papers labelled with a disparity measure of \textit{strong}, \textit{medium} or \textit{weak}, we further annotated that paper along a binary explicitness measure with the values \textit{explicit} or \textit{implicit}.
        \begin{itemize}
            \item A paper is labelled \textit{explicit} if it unambiguously names the mentioned or referenced groups, communities, or demographics that are disproportionately impacted. The annotator then records the mentioned groups, communities, or demographics.
            \item The paper is labelled \textit{implicit} if it discusses or mentions harm or impact without naming specific impacted groups, communities, or demographics and instead deals with harm in an implied, inferred, or indirect manner.
        \end{itemize}
  
\end{enumerate}

Some of the papers annotated did not fit our annotation scheme. For example, one paper focused on the social and ethical concerns of the Chinese Social Credit System~\cite{engelmann2019clear}. Although this paper fits within the broad category of addressing the ethics of algorithmic systems and their societal implications, the framing, perspective, and central subject did not fit the annotation scheme we had outlined well. The paper discussed the Chinese Social Credit system, what constitutes ``good'' and ``bad'', and the complex connection between moral behavior and external material reward. 
Although this was a compelling work, it was challenging to translate it into a traditional ``fairness'' template most papers within FAccT and AIES seem to adhere to. This is not to say that this (and the other papers) are outside the remit of AI Ethics, but rather a reflection of the limitation of our annotation scheme.

\section{Quantitative summary}
\label{sect:quant}
This section 
presents the results of our analysis of papers from FAccT and AIES across the past 4 years (2018-2021). Sub-Section~\ref{subsec:results} shows the results and how the papers were distributed along the different dimensions we surveyed, and Sub-Section~\ref{subsec:discussion} provides an in-depth discussion of the results with relation to the larger context of the paper.

\subsection{Results}
\label{subsec:results}

\begin{figure}[htbp]
    \begin{minipage}{.45\textwidth}
        \centering
        \includegraphics[width=\textwidth]{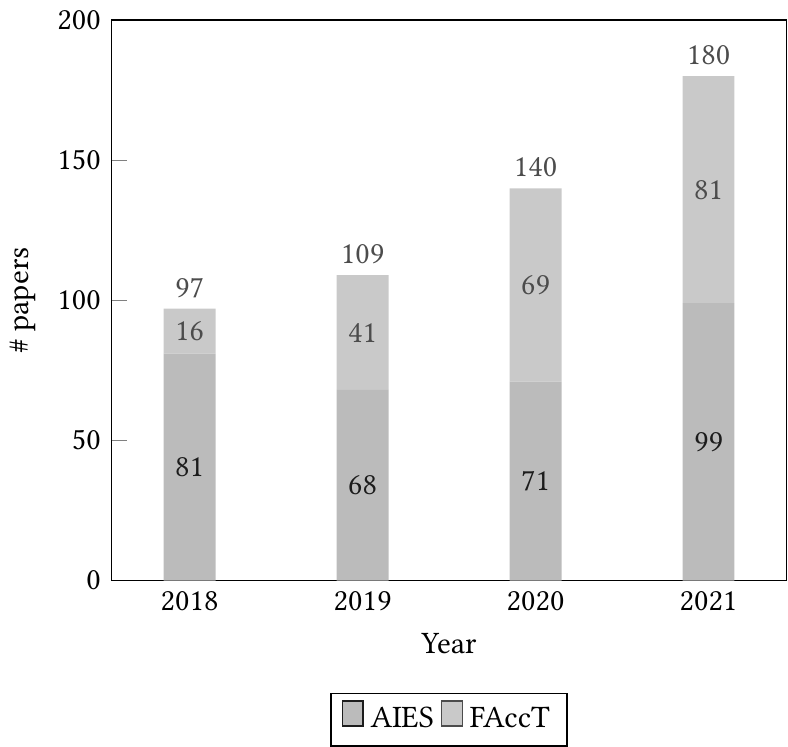}
        \captionof{figure}{Number of papers annotated by conference for each year.}
        \label{fig:number_papers_bar}
        \Description{A bar graph showing the number of papers annotated per conference per year. 97 papers for 2018 (81 AIES, 16 FAccT), 109 for 2019 (68 AIES, 41 FAccT), 140 for 2020 (71 AIES, 69 FAccT), and 180 for 2021 (99 AIES, 81 FAcct).}
    \end{minipage}%
    \hfill
    \begin{minipage}{.45\textwidth}
        \centering
        \includegraphics[width=\textwidth]{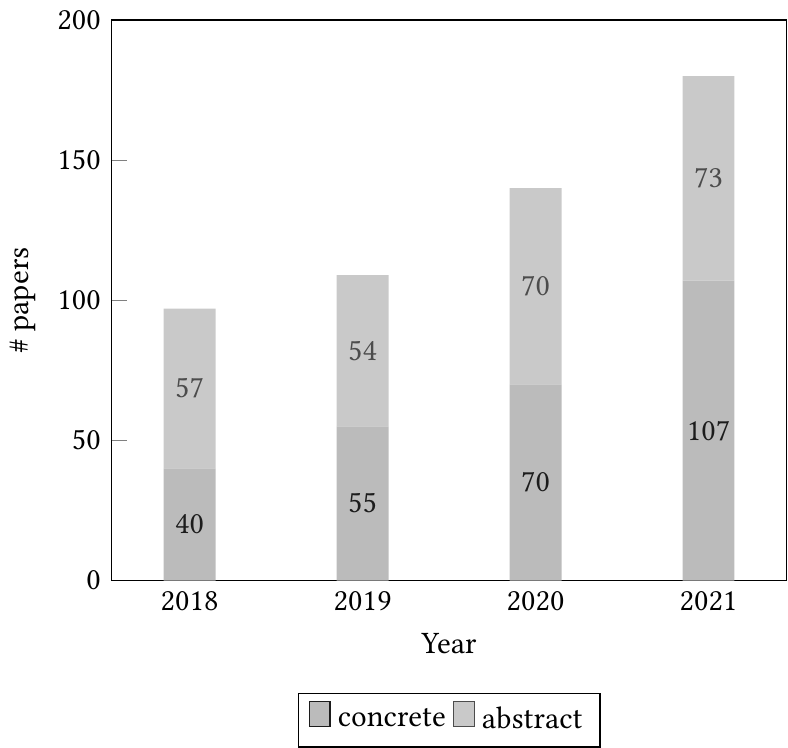}
        \captionof{figure}{Number of papers per concreteness value per year.}
        \Description{A bar graph showing the number of papers annotated as concrete and abstract per year. 40 concrete and 57 abstract for 2018, 55 concrete and 54 abstract for 2019, 70 concrete and 70 abstract for 2020, and 107 concrete and 73 abstract for 2021.}
        \label{fig:concrete_abstract_year}
    \end{minipage}%
\end{figure}

The annotated papers are distributed across the 2 conferences over the 4 years. 
As shown in Figure~\ref{fig:number_papers_bar}, the number of papers published increased each year. 
While the number of papers from AIES saw some variations, the number of papers from FAccT grew significantly and consistently over the years.

\begin{table}[htbp]
    \centering
    \caption{Percentage and number of \textit{concrete} papers per conference per year.}
    \label{tab:concrete_summary}
    \begin{tabular}{lrrrrrrrrrr}
        \toprule
        \multirow{2}{*}{\textbf{Conference}} & \multicolumn{2}{c}{\textbf{2018}} & \multicolumn{2}{c}{\textbf{2019}} & \multicolumn{2}{c}{\textbf{2020}} &
        \multicolumn{2}{c}{\textbf{2021}} &
        \multicolumn{2}{c}{\textbf{Total}} \\ 
        \cmidrule(lr){2-3}\cmidrule(lr){4-5}\cmidrule(lr){6-7}\cmidrule(lr){8-9}\cmidrule(lr){10-11}
        & \% & \# & \% & \# & \% & \# & \% & \# & \% & \#\\
        \midrule
        \textbf{AIES} & 38 & 31 & 51 & 35 & 32 & 23 & 64 & 63 & 47.6 & 152\\
        \textbf{FAccT} & 56 & 9 & 49 & 20 & 68 & 47 & 54 & 44 & 58 & 120\\
        \textbf{Total} & 41 & 40 & 50 & 55 & 50 & 70 & 59 & 107 & 52 & 272\\
    \bottomrule
    \end{tabular}
\end{table}


The number of \textit{concrete} and \textit{abstract} papers remained relatively proportional for the first 3 years while the last year (2021) saw more (107) concrete papers compared to abstract (73) as shown in Figure~\ref{fig:concrete_abstract_year}. Table~\ref{tab:concrete_summary} details the distribution for each of the 2 conferences, showing both the absolute number and proportion of \textit{concrete} papers in each conference for each year. The proportion of concrete papers grew slightly over the 4 considered years, from 41\% in 2018 to 59\% in 2021. Over this period, FAccT accepted proportionally more \textit{concrete} papers than AIES.

\begin{figure}[htbp]
    \centering
    \includegraphics[width=\columnwidth]{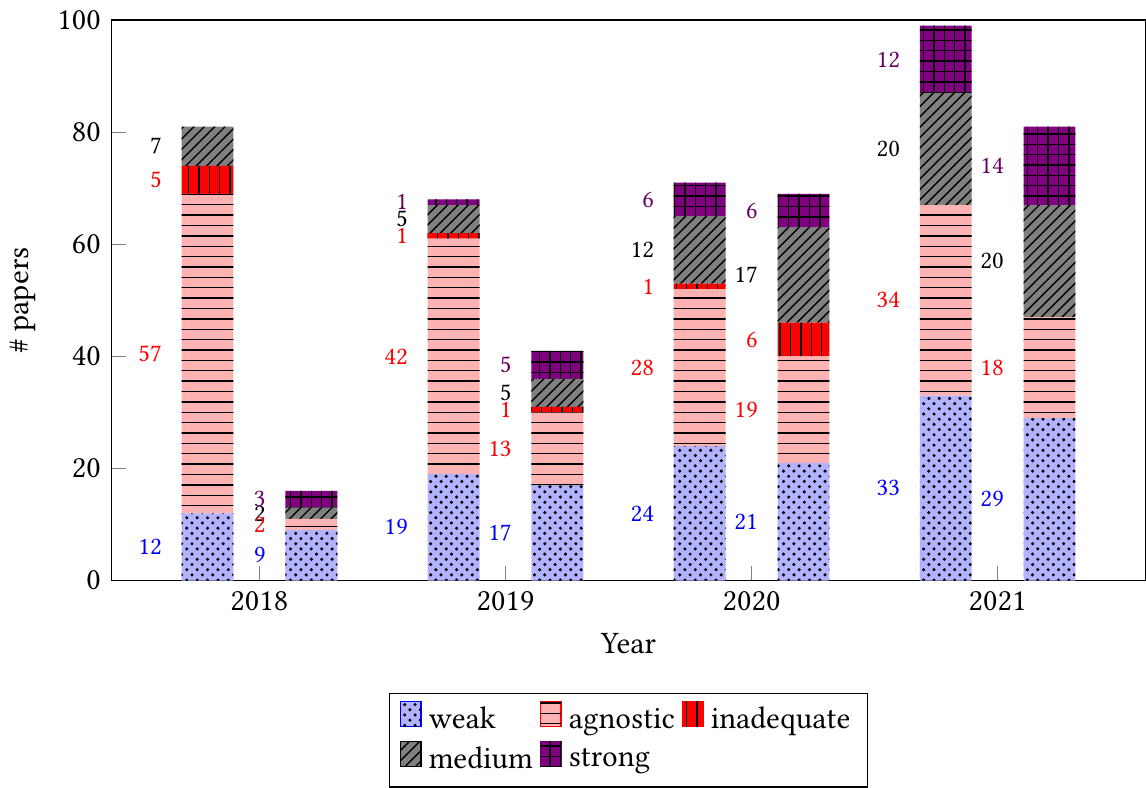}
    \caption{Number of papers for each disparity measure value per year for AIES (left), and FAccT (right).}
    \label{fig:disparity}
    \Description{A bar graph showing the number of papers annotated with each disparity measure value per conference per year. Agnostic and weak paper represent a majority of papers for both conferences for both years.}
\end{figure}

\begin{table}[htbp]
    \centering
    \caption{Percentage and number of papers with agnostic position on disparate impact per conference and per year.}
    \begin{tabular}{lrrrrrrrrrr}
        \toprule
        \multirow{2}{*}{\textbf{Conference}} & \multicolumn{2}{c}{\textbf{2018}} & \multicolumn{2}{c}{\textbf{2019}} & \multicolumn{2}{c}{\textbf{2020}} &
        \multicolumn{2}{c}{\textbf{2021}} &
        \multicolumn{2}{c}{\textbf{Total}} \\ 
        \cmidrule(lr){2-3}\cmidrule(lr){4-5}\cmidrule(lr){6-7}\cmidrule(lr){8-9}\cmidrule(lr){10-11}
        & \% & \# & \% & \# & \% & \# & \% & \# & \% & \#\\
        \midrule
        \textbf{AIES} & 70 & 57 & 61 & 42 & 39 & 28 & 34 & 34 & 50 & 161\\
        \textbf{FAccT} & 12 & 2 & 32 & 13 & 28 & 19 & 22 & 18 & 25 & 52\\
        \bottomrule
    \end{tabular}
    \label{tab:agnostic_per_year_per_conf}
\end{table}

Figure~\ref{fig:disparity} and Table~\ref{tab:agnostic_per_year_per_conf} present results related to the disparity measure. As Figure~\ref{fig:disparity} shows, overall, the \textit{agnostic}, \textit{weak}, and \textit{medium} categories represent most of the papers, with 40\%, 31\%, and 17\% of all papers respectively. This is particularly true in the earlier years. It appears we may be seeing the beginning of a trend towards increased numbers of \textit{medium} and \textit{strong} papers in both venues as the proportion of \textit{agnostic} papers that do not mention any disparate impact of algorithmic systems declines. Indeed, in 2018, the \textit{medium} and \textit{strong} categories represented 12\% of papers, and the \textit{agnostic} papers 61\%; while in 2021, \textit{medium} and \textit{strong} covered 37\% of papers and \textit{agnostic} decreased to 29\%. Table~\ref{tab:agnostic_per_year_per_conf} further delves into the \textit{agnostic} category and shows, for each conference and each year, the number and proportion of these papers. For AIES a significant proportion of papers (49\% overall) fall into the \textit{agnostic} category, but this proportion has decreased over the years.

\begin{figure}[htbp]
    \begin{minipage}{.45\textwidth}
        \centering
        \includegraphics[width=\textwidth]{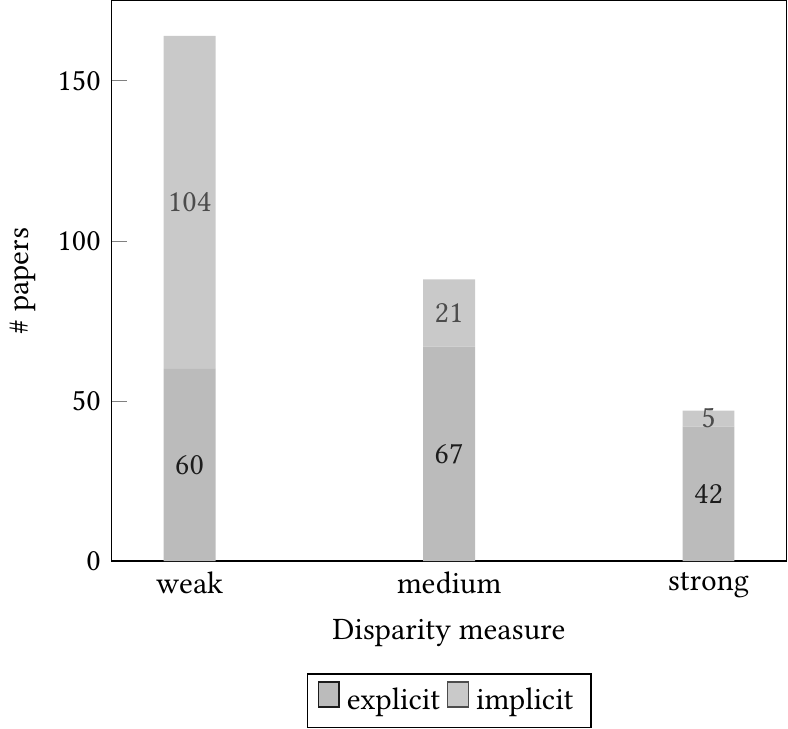}
        \captionof{figure}{Number of paper per explicitness per disparity measure value.}
        \label{fig:implicit_explicit}
        \Description{A bar graph showing the number of papers annotated explicit and implicit per disparity measure value. 60 explicit, 104 implicit for weak papers, 67 explicit, 21 implicit for medium papers, 42 explicit and 5 implicit for strong papers.}
    \end{minipage}%
    \hfill
    \begin{minipage}{.45\textwidth}
        \centering
        \includegraphics[width=\textwidth]{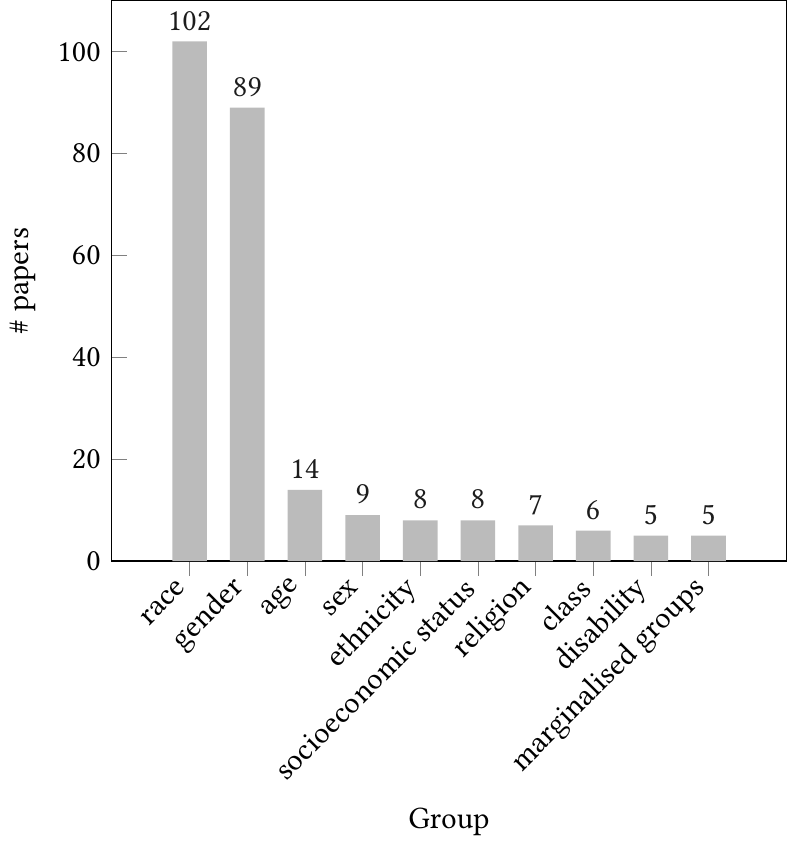}
        \captionof{figure}{10 most mentioned impacted groups.}
        \label{fig:groups}
        \Description{A bar graph showing the 10 most mentioned impacted groups in the surveyed paper and the number of papers mentioning them: race (102), gender (89), age (14), sex (9), ethnicity (8), socioeconomic status (8), religion (7), class (6), disability (5), marginalised groups (5).}
    \end{minipage}%
\end{figure}

Next, we further break down the explicitness measure across the three disparity categories: 
\textit{weak}, \textit{medium}, and \textit{strong}. As can be seen in Figure~\ref{fig:implicit_explicit},  
most papers (62\%) in the \textit{weak} category use \textit{implicit} disparity measures, while those in the \textit{medium} and \textit{strong} category tend to be more \textit{explicit} in their discussions of algorithmic harm (76\% and 89\% respectively), unambiguously naming the communities and demographics that are the focus of study. Of those, Figure~\ref{fig:groups} shows the 10 groups that were most mentioned or discussed in the annotated papers, and for each group the number of papers that mentioned it. Race and gender are, by far, the two most mentioned dimensions in impacted groups.

\subsection{Discussion of the results}
\label{subsec:discussion}

The results of this survey give insight both into the current state and potential directions of AI Ethics scholarship at each conference.
FAccT began as a much smaller venue, with a particular 
focus that framed algorithmic systems as socio-technical systems, and attention to 
interdisciplinarity. 
This focus was reflected in the papers in the inaugural conference (see Table~\ref{tab:concrete_summary}) where the majority of the (albeit small number of) papers were \textit{concrete}. Furthermore, very few of those papers were \textit{agnostic} and thus very few did not engage in some way with how the socio-technical systems in question 
relate to some form of oppression or marginalization. Papers published in FAccT showed consistency across the years in (at least) naming particular groups that have been negatively impacted by socio-technical systems, as can be seen from the non-\textit{agnostic} counts in Figure.~\ref{fig:disparity}.

Though AIES also has always had a multidisciplinary focus, most of the papers presented at the inaugural conference had a much more abstract focus, and the majority of those papers had an agnostic position. This abstract and agnostic focus perhaps points to less involvement from works 
that center the experiences of marginalized communities when examining 
AI systems. 
Many of the papers failed to engage with any intersections between the chosen topic within AI ethics in a concrete way, i.e., in a way that acknowledged the potential intersections between their system or approach and some form of oppression or marginalization. Furthermore, a relatively large portion of the papers during that inaugural conference were annotated as ``inadequate'' in some way, telling a troubling story of the initial wave of papers approaching this topic. Despite the troubling beginning to the proportion of inadequate papers, we did see an improvement, after the inaugural year. 
The more recent shift towards a smaller proportion of \textit{agnostic} papers and a relatively higher proportion of both \textit{strong} and \textit{medium} papers hopefully points to an increasing recognition among the community of the importance of work that considers the interactions between the socio-technical system in question, 
impacts on marginalized communities, and systems of oppression that result in this marginalized status.

Overall, the conferences appear to have more similar proceedings in terms of concreteness (Table~\ref{tab:concrete_summary}) and non-\textit{agnostic} (Table~\ref{tab:agnostic_per_year_per_conf}) positions in the most recent year (2021), than when they began. 
Nonetheless, 
most papers consistently were either \textit{agnostic} or \textit{weak} (Figure.~\ref{fig:disparity}) which indicates that much of the work is lacking in historical and ongoing contextualization of 
socio-technical systems. 
We also see that papers with \textit{strong} disparity measure tend to explicitly reference impacted groups \textit{the most} (89\%). This tendency decreased with \textit{medium} papers (76\%), while only a minority of \textit{weak} papers were \textit{explicit} (37\%) in their naming of particular groups harmed or disparately impacted  during their discussions (Figure.~\ref{fig:implicit_explicit}). This is a telling even if somewhat expected result --- it is difficult for papers to not explicitly name impacted groups when the core focus 
centers around exposing or tackling disparate impacts. With \textit{weak} papers, it is easier, and perhaps more likely, to see examples of work approaching disparate impact that does not explicitly name who is impacted and how they are impacted. 
With the majority of papers we surveyed being \textit{agnostic} or \textit{weak}, and thus either implicitly mentioning harms or not addressing them at all, we see an overall inadequacy of AI Ethics scholarship in that it lacks 
engagement with perspectives and histories that are not a part of the standard, or the Human~\cite{Wynter03}.

The number of papers that we annotated as inadequate is relatively low compared to the total number of papers appearing at both venues, however their presence itself warrants further discussion. 
Most of those papers appeared in 2018 and 2020 (with AIES having the only papers rated in such a way during 2018 and FAccT having the majority of papers falling in that category in 2020). This small subset of papers was varied in the ways in which the papers were inadequate–--some failed to consider potentially problematic uses of their systems given the context in which they discussed them, while others seemed to positively engage with perspectives that could be viewed as anti-equity. One example of the former is presented in Rodolfa et al.~\cite{rodolfa20recidivism}. The authors present a system they developed in collaboration with the Los Angeles City’s attorney office that uses predictive modeling to prepare social services resources for those likely to be re-arrested (with a focus on those with prior interactions and arrests with the police). Understandably, the authors want to find a better use of these predictive algorithms than re-arrest and the authors include a healthy discussion on equity in the predictive models. However, the considerations of how these predictive algorithms fit into the institution-person interaction, especially from the perspective of the person who has been re-arrested (or may be in the future) are extremely limited. At its core, this inadequacy stems from a shallow consideration of ``bias'' that does not more readily approach the racial (and interlocking class and gender) systems that undergird an individual's interactions with these criminal justice institutions. Without such considerations, the 
predictive modeling into 
existing racist, capitalist, and unjust institutions such as the US criminal justice system \cite{muhammadblackness,alexanderjimcrow}, risks becoming 
anti-Black. 

Another paper that fell under the \textit{inadequate} category discussed affirmative action.  
The paper by Kannan et al.~\cite{kannan2019affirmativeaction} uses computational modeling to explore aspects of downstream effects of affirmative action policies in undergraduate college admissions. In the limitations section of the paper the authors note that their approach does not account for ``past inequity'' and ``...does not attempt to address or correct the historical forces...''. While the goal of finding a new way to understand effects of these policies is understandable, failing to consider these critical factors  amounts to `historyless' 
exploration, giving the illusion that current policies and realities can be severed from their unjust historical pasts. Such work is likely be used as way to bolster current systems of racism~\cite{bonilla2015structureracism}, and 
is 
antithetical to 
movements that have led to more progressive policy such as affirmative action (e.g., see~\cite{mlk1968}). The very assumption that past inequity is irrelevant to current affirmative action policies risks such papers becoming anti-equity. 

\section{Qualitative analysis: ProPublica's COMPAS as a case study}
\label{sect:qual}
The ProPublica  COMPAS study~\cite{angwin2016machine} was one of the most recurring citations in the papers we studied (e.g., \cite{kim2019multiaccuracy, matthews2019righconfront, gilbert2019epistemic, mcnamara2019equalized, mishler2019modelingrisk, srivastava2018composablebias, vasconcelos2019epistemological, kalyanakrishnan2018aiindia, goel2018mlconvex, couldechova2018child, buolamwini2018gender, taskesen21compas, fogliato21compas, krafft21compas}). We found that many of the papers failed to engage beyond a shallow mention of the paper or the associated dataset. Instead, these papers frequently focused on ways to improve fairness in assessing likelihood to re-offend or, more generally, ways to increase fairness within current incarceration processes, with authors finding ways within their specific approaches to train a system that produced fairer results.

Most of these papers 
failed to address what it means to re-offend within a society (particularly in the context of the US criminal justice system) that deems some people more likely to offend than others (e.g., due to their race, sex, gender, class status, and intersections thereof or, relatedly, due to their ``governability''~\cite{quan2017demliving}). Many of these papers also did not address 
questions such as how the dataset is sourced and curated
, what was used as input data to build the algorithm, how historical and current stereotypes are encoded~\cite{benjamin2019race,birhane2021large}, and whether improving recidivism algorithms 
results in a net harm to marginalized groups. 

As an example of approaches that can be critiqued, Taskesen et al.~\cite{taskesen21compas} describe a framework for assessing probabilistic fairness for classifiers. The authors note that we have recently seen ``data and algorithms'' become ``an integrative part of the human society'' and use the COMPAS dataset to provide an exemplar of their proposed statistical test of fairness in use. While, commendably, the paper does focus on fairness and tests it on the real-world dataset, it critically lacks in its contextualization of the very systems that help define the problem space and environment in which the classifiers (that they hope to assess for fairness) operate. This lack of context obscures a very important question of how probabilistic fairness interacts with the historical and ongoing dehumanizing processes; those processes that help define who is functionally deemed a part of ``human society''. The lack of a historical context also results in incorrectly labeling \textit{the use of} datasets and algorithms as a new practice. Although, our use of digital systems to represent, extend, and modify data and algorithms has certainly changed, what we do with digital systems now is a continuation of decades old \textit{computing} systems. Though the paper does conclude with a note on problematic potential of auditing tools falling in the hands of ``vicious'' inspectors, such an individual focus fails to consider that such audits occur within a context of systems that impose hierarchies determined by hegemonic knowledge. That is, the use of the word ``vicious'' obscures the mundaneness of the violence enacted by these systems and thus the normalcy of an auditor’s behavior when they use such a tool in a way that results in, for example, racial violence.

It is unlikely to be impactful to consider the COMPAS dataset as being fairer in the ratings of likelihood of re-offense if we fail to explicitly take into account why Black communities are more likely to be regarded as criminal, which may be due to their physical or social proximity to other previous ``offenders'', or the spaces that have historically excluded them, for example, racially and class segregated neighborhoods~\footnote{See~\cite{gilmore2017abolition} for a related discussion on the problem of \textit{innocence}, as well as~\cite{weheliye2014habeas} for a discussion of relations between Blackness and Law.}. These perspectives fail to adequately address how the institutional contexts in which those systems are applied (i.e., as socio-technical systems) bring with them a socio-cultural milieu that defines not only how those systems are implemented and integrated, but also how those institutions themselves adapt to any given deployment or integration of technology. For example, they often fail to critically address the anti-blackness that undergirds the socio-technical system they hope to improve (e.g., see~\cite{dancy21blackness} for a discussion of the need to move beyond notions of representation and bias in the development of AI systems). More importantly, the fixation on improving such algorithms, by tweaking them to make them fairer or more accurate, contributes to maintaining punitory systems, rendering reformative questions such as ``what mechanisms would be put in place to help the convicted integrate back into society after release?'' out of the purview of AI ethics. These nuances, perspectives, and important backgrounds are lost as we approach ethics, fairness, and related concepts as solely abstract mathematical formulations or philosophical musings. An approach cannot adequately address these concerns unless it engages with the socio-cultural institutions that 
ultimately enact it.

Having said that, acknowledgement of systems of oppression in the context of COMPAS is not entirely absent. Fogliato et al~\cite{fogliato21compas} and Krafft et al.~\cite{krafft21compas}, for example, provide examples of more directly engaging with issues related to systems of oppression and ideas of power, even if they do so with different aims. Fogliato et al~\cite{fogliato21compas} present their investigation of biases (particularly racial) in arrest data that might be used by risk assessment instruments (such as COMPAS). Though the paper falls short in some areas of addressing anti-blackness as a concept directly~\footnote{For example, in their analysis of the city of Memphis, the authors could have noted and traced historical anti-blackness~\cite{wells1893lynch} and how surveillance~\cite{jefferson2020digitizepunish} and violence enacted by policing systems relate.}, they did discuss their work in the context of related concepts, such as victim devaluing and police response to criminal acts in racially segregated (predominantly Black) areas. Krafft et al.~\cite{krafft21compas} describe the development and use of an AI auditing toolkit that can be used as a development and deployment decision tool. Deviating from typical practice, during the design of this toolkit, the authors explicitly engaged with a wider audience, including community organizers and those with the lived experience of marginalization; this community orientation  helps to center the issues faced by those communities. Though both of these papers have a different stated aim than many of the other papers we reviewed that discuss and use COMPAS and the Propublica study, we 
argue that work in this space should 
center the participation, perspective, and input of the communities affected by the systems they wish to make more ``fair''. Such work should also be contextualized in 
historical and ongoing oppressive systems and institutions that result in disproportionate harm on those marginalized and dehumanized communities.

\section{Conclusion --- Missing the forest for the trees}
\label{sect:conclusion}

Even though AI Ethics is a fast growing and broadly construed  field of enquiry, its pace is no match for the rate at which algorithmic systems are being developed and integrated into every possible corner of society. 
Thus, the field holds a crucial place in ensuring that 
algorithmic systems are just and equitable; in bringing to light algorithmic failures, whom they fail; as well as in holding responsible bodies accountable. 
If AI Ethics is to protect the welfare and well-being of the most negatively impacted stakeholders, it needs to be guided by the lived experiences and perspective of such stakeholders. The field also needs to treat AI Ethics as non-divorcible from day-to-day life and something that can't emerge in a historical, social, cultural, and contextual vacuum. 

A review of the field via analysis of 
the two most prominent conferences shows that there is a great tendency for abstract discussion of such topics devoid of structural and social factors and specific and potential harms. While there is a welcome reduction in percentage of papers that are agnostic to specific forms of oppression or marginalization, we also found that those papers that are agnostic to or only \textit{weakly} address those contexts remain the majority of the papers in each conference. As discussed in section \ref{sect:qual}, we also found that many papers (in this case those citing the ProPublica COMPAS study) failed to directly center structural (social) systems, as well as \textit{institutions} that are used to enact those systems. The lack of a critical perspective on these institutions resulted in what we saw as an inadequate treatment of issues relating to ethics. 
Furthermore, some papers judged as inadequate (as discussed in section \ref{subsec:discussion}) bordered on forms of anti-equity given the topics approached, the institutions involved, and the absence of a critical perspective on those institutions. 

Given that the most marginalized in society are the most impacted when algorithmic systems fail, we contend that all AI ethics work, from research, to policy, to governance should pay attention to structural factors and actual existing harms. In this paper, we have argued that given oppressive social structures, power asymmetries, and the uneven harm and benefit distributions work in AI Ethics should pay more attention to these broad social, historical, and structural factors in order to bring about actual change that benefits the least powerful in society. 

\begin{acks}
    This work was supported, in part, by Science Foundation Ireland grant 13/RC/2094\_P2, and co-funded under the European Regional Development Fund through the Southern \& Eastern Regional Operational Programme to Lero - the Science Foundation Ireland Research Centre for Software (www.lero.ie). This work was also partially supported by the National Science Foundation under grant No. 2218226. 
\end{acks}

\bibliographystyle{ACM-Reference-Format}
\bibliography{FAccT}

\appendix

\end{document}